\documentclass{iau}
\usepackage{graphicx}

\newcommand{\HI}{H\,{\sc i}}
\newcommand{\U}[1]{\ensuremath{\mathrm{#1}}}
\newcommand{\kms}{\U{km\,s^{-1}}}
\newcommand{\Ht}{H$_{2}$}

\title[The Chaotic THINGS survey] 
{Probing the star--formation modes in merging galaxies}

\author[P.-A.  Duc  et al.]   
{P.--A. Duc$^1$,
 P.-E. Belles$^1$$^,$$^2$,
 E. Brinks$^2$
\and F. Bournaud$^1$  
 }

\affiliation{$^1$AIM Paris Saclay, F-91191 Gif sur Yvette cedex;     email: {\tt paduc@cea.fr} \\[\affilskip]
$^2$ University of Hertfordshire, College Lane, Hatfield AL10 9AB, UK }

\pubyear{2012}
\volume{292}  
\pagerange{119--126}
\jname{Molecular Gas, Dust, and Star Formation in Galaxies}
\editors{A.C. Editor, B.D. Editor \& C.E. Editor, eds.}
\editors{T. Wong \& J. Ott, eds.}

\begin{document}

\maketitle

\begin{abstract}
Merging systems at low redshift provide the unique opportunity to study the processes related to star formation in a variety of environments that presumably resemble those seen at higher redshifts. Previous studies of distant starbursting galaxies suggest that stars are born in turbulent gas, with a higher efficiency than in MW-like spirals. We have investigated in detail the turbulent-driven regime of star-formation in nearby colliding galaxies combining  high resolution VLA  B array \HI~ maps and  UV {\it GALEX} observations. With these data, we could check predictions  of our state-of-the-art simulations of mergers, such as the global sharp increase of the fraction of dense gas, as traced by the SFR, with respect to the diffuse gas traced by \HI\   during the merging stage, following the increased velocity dispersion of the gas. We present here initial results obtained studying the SFR-\HI~ relation at 4.5~kpc resolution. We determined    SFR/\HI~mass ratios that are higher in the external regions of mergers than in the outskirts of isolated spirals, though both environments are \HI~dominated. SFR/\HI~ increases towards the central regions following the decrease of the atomic gas fraction and possibly the increased star--formation efficiency.
These results need to be checked with a larger sample of systems and on smaller spatial scales. This is the goal of the on-going Chaotic THINGS project that ultimately will allow us to  determine why starbursting galaxies deviate from the Kennicutt-Schmidt relation between SFR density and gas surface density.

\end{abstract}

\firstsection 
\section{Introduction}

\begin{figure}[t]
\begin{center}
 \includegraphics[width=0.85\textwidth]{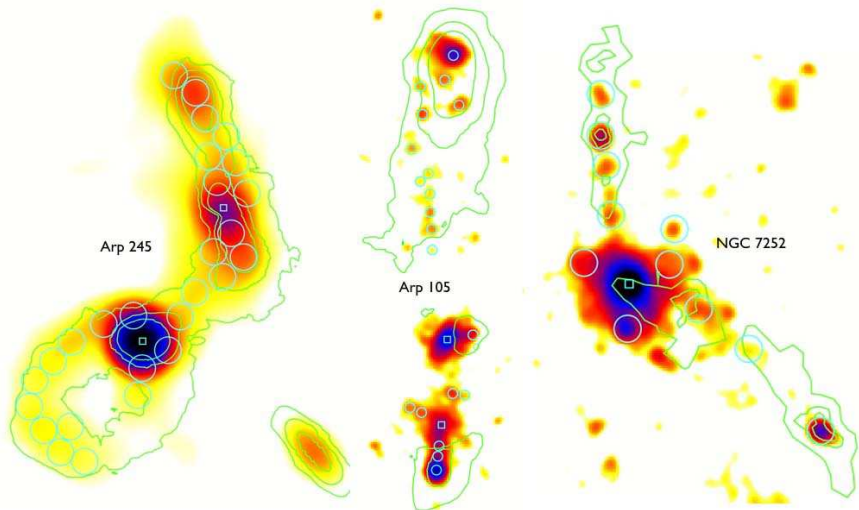} 
 \caption{ \HI\  contours superimposed on the UV maps  of three nearby interacting systems. The regions for which a region-to-region analysis of the SFR--\HI\ relation are indicated with  circles.}
   \label{fig1}
\end{center}
\end{figure}

The role of starbursting mergers versus quiescent evolution in the star formation (SF) budget of galaxies is one of the most debated questions in galaxy formation, both theoretically \cite[(Dekel et al., 2009]{Dekel09}) and observationally \cite[(Daddi et al., 2010]{Daddi10}). 
The debate has intensified in recent years after it was realized that (i) high-redshift disk galaxies can drive an active phase of star formation just by internal processes, as they are very gas--rich and (ii) their large scale interstellar medium (ISM) is highly turbulent, with turbulent speeds of several tens of \kms
  \cite[(F{\"o}rster Schreiber et al.,  2006]{FS06}). 

 ISM turbulence plays a fundamental role in driving and regulating star formation in any galaxy, from disks to mergers. On the one hand, turbulent flows can locally compress the gas, and form over--densities that subsequently cool and collapse into star--forming clouds. On the other hand, turbulence can support clouds against rapid collapse, or even disrupt local over--densities in the ISM before they have time to form stars, hence regulating star formation \cite[(Elmegreen et al., 2002]{Elmegreen02}). Extreme turbulence in high--velocity shocks may even suppress star--formation \cite[(Appleton et al. 2006)]{Appleton06}.  Differences in the turbulent nature of the ISM in galaxies could thus lead to significantly different SF regimes, and this could be particularly important in colliding and merging galaxies in general, and in high--redshift galaxies in particular.

Star formation on galactic scales can in general be seen as a two--step process: (1) formation of dense molecular gas clouds in the large scale ISM -- a first step driven by gravitation and cooling processes  and (2) star formation in the dense cores of these molecular clouds. Whereas the conversion of dense molecular gas into stars may follow a universal efficiency, the first phase, namely the formation of molecular complexes from large-scale atomic reservoirs, is a less studied process.

 While surveys of high-redshift disks show that they are highly turbulent, they do not resolve their gas properties down to the scale of star--forming complexes. ALMA will map the densest phases of the molecular gas, thus probing the final steps of star formation in dense clouds, but not the formation of these dense clouds from lower--density, atomic gas. 
Thus, surveys of the atomic gas are indispensable in order to address the following questions: how efficient is star formation in a more turbulent ISM than in quiescent (nearby) spirals? 
Are the strongest starbursts at any redshift (linked to mergers) triggered mostly by ISM turbulence, or by smaller scale processes (star formation inside dense clouds) or larger scale ones (global gas inflows)?

In the absence in the nearby Universe of the equivalent of the distant gas--dominated and highly turbulent disks, galaxy mergers offer unique local laboratories to probe ISM physics and SF processes in such turbulent environments.  Indeed, their gas velocity dispersion can be several tens of \kms\, just like in high-redshift galaxies. 

 Increased ISM turbulence in interacting galaxies can now be modeled in detail in simulations with modern grid--based hydro--codes (AMR) with spatial resolutions up to one parsec  \cite[(Teyssier et al. 2010)]{Teyssier10}.
These simulations that directly resolve star--forming clouds  show that this turbulence--driven star formation can trigger starbursts simply by changing  the distribution of the gas between low--density phases and dense star--forming phases. \cite[Teyssier et al. (2010)]{Teyssier10} thus propose that the recently observed deviation of mergers from the Kennicutt--Schmidt (KS) relation between the SFR and gas surface density  \cite[(Daddi et al., 2010]{Daddi10}) can be explained mostly by turbulence.
The change of the Probability Distribution Function (PDF) of the ISM density  due to the increased turbulence in mergers is illustrated in \cite[Bournaud (2011)]{Bournaud11}. 
Observationally, it translates to a local increase of the Star Formation Rate  to HI mass ratio during the merger.
A theoretical modeling of the change of the PDF shape and its impact on the  KS relation has recently been proposed by \cite[Renaud et al.  (2012)]{Renaud12}.

We have checked the predictions of the simulations and models studying the spatially resolved SFR--\HI~ relation in a sample of nearby interacting systems.


\begin{figure}[b]
\begin{center}
 \includegraphics[width=\textwidth]{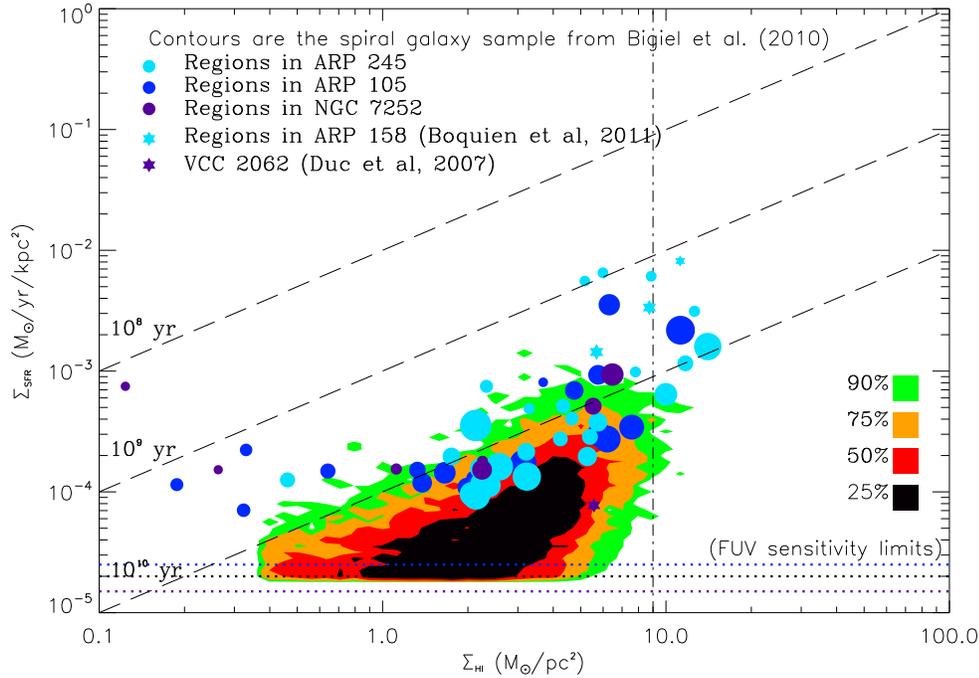} 
 \caption{$\Sigma_{SFR} - \Sigma_{HI}$ density relation for our sample of star forming regions in mergers.  All fluxes were corrected for internal and Galactic extinction. The size of the symbols is directly proportional to the distance of the star forming region to the central galaxy. For comparison, the data obtained by the pixel-to-pixel study at 750~pc resolution of  the outer discs of spiral galaxies by \cite[Bigiel et al. (2010)]{Bigiel10}   are plotted, together with 
additional data on mergers taken from the literature:  Arp\,158 and VCC~2062.  The dashed lines represent lines of constant star formation efficiency (or constant depletion time). The vertical dotted line represents the saturation value of \HI, as measured by \cite[Bigiel et al. (2010)]{Bigiel10}   from a sample of spiral galaxies.}
   \label{fig2}
\end{center}
\end{figure}

\section{Spatial evolution of the  \HI - SFR relation in sample of nearby colliding systems}

High resolution maps of the \HI~gas were obtained with the JVLA  in B configuration for a sample of 7 nearby colliding systems,  as part of the Chaotic THINGS survey. We present here initial results for 3 systems: 2 interacting galaxies: Arp~105 and  Arp~245 and the  well known advanced  merger NGC~7252
(see Fig~\ref{fig1}). According to their integrated properties and location on the KS relation, they are observed during  a starburst phase.

Their external regions, in particular their tidal tails, turn out to be  gas-rich, with the \HI\  dominating by far the  baryonic budget 
whereas, for two galaxies,  the inner regions are \HI--poor. 
The  Star Formation Rate (determined from UV emission from {\it GALEX})  and  \HI~mass per unit area were compared at the 4.5\,kpc-scale. 
A tight spatial  correlation  between the UV and \HI~emission was found.
Fig~\ref{fig2} plots the individual extracted regions on the SFR -- \HI~surface density  plane.  For comparison, the results of the pixel-to-pixel analysis of external regions of isolated spirals, observed as part of the THINGS survey \cite[(Bigiel et al., 2010)]{Bigiel10},  are  also shown. 
The SFR/\HI~mass ratio in the external regions of mergers, a proxy of the Star-Formation Efficiency (SFE)  in that environment,  
 was found to be higher than that found in \HI--dominated regions of isolated systems, like the outer discs of spiral galaxies.
Besides, it only weakly varies with   gas density, just like  the SFE of  the internal disks of isolated spirals, characterized by a high molecular gas fraction \cite[(Bigiel et al., 2008)]{Bigiel08}. In other words, the outskirts of mergers, though they are formally \HI--dominated, show a regime of Star-Formation more characteristic of the one that prevails  in the  \Ht--dominated regions of isolated spirals.  

As shown on Fig~\ref{fig2}, the SFR/\HI~mass ratio seems to increase towards the inner most regions, partly due to the decrease of the \HI~column density (following the transition of the atomic gas to a molecular phase), or possibly due to an increase of the SFE, here defined as the ratio between the SFR to the molecular gas mass.

\section{Evolution of the Star Formation Efficiency}


A proper study of the spatial  (in-out) and temporal  (along the merger sequence) evolution of the SFE would require to  map  the molecular gas of several systems over their full \HI\ extent. This is  now achievable with the IRAM and ALMA facilities. 
A few pointed CO  observations towards  tidal tails  already  exist, including for the systems studied here. These external regions of mergers  have a molecular gas fraction of typically 30 to  40\%, i.e., less than that measured in the disks of isolated spirals like the MW, while their star--forming regions  have similar SFEs  \cite[(Braine et al., 2001)]{Braine01}. This is consistent with the results obtained analyzing the \HI~data:  the SF regime in the external parts of mergers is closer to what prevails in the inner disks of spirals than in the outer ones which is very inefficient at forming stars. 
 There are some exceptions though: in shock regions such as the X--ray ridge of  Stephan's Quintet \cite[(Guillard et al., 2012)]{Guillard12}, or in the collisional debris of ancient mergers, such as NGC~4694 / VCC~2062 (Lisenfeld et al., 2013,  in prep.), low SFE have been measured.
Arp~158 is one of the very few systems where \HI, CO, UV/IR data are available over the whole system. Interestingly, \cite[Boquien et al. (2011)]{Boquien11} found enhanced  SFE towards its merging nuclei, while the SFE in the other regions, in particular in the tidal tails, is constant (see Fig.~\ref{fig2}). So the two branches of the  KS relation (for starbursts and normal spirals)  are observed there within a single system. 

Studying the SFR-gas relation {\it at the kpc-scale} for a larger number of systems is most needed to further  investigate the origin of the global enhanced SFE found in nearby and distant starbursts.


\end{document}